\documentclass[12pt]{article}
\usepackage{epsfig,amsfonts,amsthm}
\usepackage{amsmath,amssymb,mathtools}
\newcommand{\be}{\begin{equation}}
\newcommand{\ee}{\end{equation}}
\newcommand{\bea}{\begin{eqnarray}}
\newcommand{\eea}{\end{eqnarray}}
\newcommand{\bs}{\begin{subequations}}
\newcommand{\es}{\end{subequations}}

\newcommand{\f}{\phi}
\newcommand{\fd}{\phi^\dagger}

\newcommand{\triplet}[3]{ \left(\! \begin{array}{c}#1 \\ #2 \\ #3 \end{array}\!\right) }

\newcommand{\lr}[1]{ \langle #1 \rangle}
\newcommand{\Z}{\mathbb{Z}}

\def\lsim{\mathrel{\rlap{\lower4pt\hbox{\hskip1pt$\sim$}}
    \raise1pt\hbox{$<$}}}         
\def\gsim{\mathrel{\rlap{\lower4pt\hbox{\hskip1pt$\sim$}}
    \raise1pt\hbox{$>$}}}         

\title{
{\normalsize \hfill CFTP/15-012} \\*[7mm]
Group-theoretic restrictions on generation of $CP$-violation in multi-Higgs-doublet models}
\author{G.~C.~Branco\footnote{E-mail: gbranco@tecnico.ulisboa.pt}\;\;~and 
I.~P.~Ivanov\footnote{E-mail: igor.ivanov@tecnico.ulisboa.pt}  
\\
  {\small CFTP, Departamento de F\'{\i}sica,
Instituto Superior T\'{e}cnico, Universidade de Lisboa,}\\ 
{\small Avenida Rovisco Pais 1, 1049 Lisboa, Portugal}\\
}

\begin{document}
\maketitle

\bigskip
\begin{abstract}
It has been known since decades that imposing a symmetry group $G$ on the scalar sector 
of multi-Higgs-doublet models has consequences for $CP$-violation.
In all examples of two- and three-Higgs-doublet models equipped with symmetries, 
one observes the following intriguing property:
if $G$ prevents explicit $CP$-violation (CPV), at least in the neutral Higgs sector, then it also prevents spontaneous CPV,
and if $G$ allows explicit CPV, then it allows for spontaneous CPV.
One is led to conjecture that this is a general phenomenon.
In this paper, we prove this conjecture for any rephasing symmetry group $G$ and any number of doublets.
\end{abstract}

\section{Introduction}

There are two distinct ways of introducing $CP$ violation (CPV) in unified gauge theories, namely,
breaking $CP$ at the Lagrangian level or breaking it spontaneously. 
At present, there is clear evidence that the CKM matrix is complex \cite{RPP2014}, 
even if one allows \cite{Botella:2005fc} for the presence of New Physics beyond the Standard
Model (SM). Therefore, in viable models of spontaneous CPV the vacuum phase(s) have to be 
capable of generating a complex CKM matrix. 
There are models which fulfill this requirement, the simplest ones \cite{Bento:1990wv} 
involve the introduction of at least one vector-like quark.

The idea of spontaneous $CP$ violation was originally suggested by T.~D.~Lee in the framework of a 
two-Higgs-doublets model (2HDM) \cite{TDLee}, with no extra symmetry introduced in the Lagrangian, apart from the gauge symmetry. 
This original model has the disadvantage of leading to too large scalar mediated Flavour-Changing-Neutral-Currents (FCNC).
Large FCNC can be naturally eliminated by imposing natural flavour conservation (NFC) \cite{NFC},
but, with two Higgs doublets, one then loses the possibility of $CP$ violation in the scalar sector.
It was later pointed out that, within the three-Higgs-doublet model (3HDM), NFC
is compatible with $CP$ violation in the Higgs sector, either explicit \cite{Weinberg} or spontaneous \cite{Branco}.
A short review of the early results can be found in \cite{Branco-1985}, for a more in-depth treatment
of $CP$ violation in multi-Higgs models see \cite{CPV}.

Already these first examples hint at an intriguing interplay between a horizontal symmetry group $G$ 
and the two forms of CPV in the scalar sector.
In 2HDM, the imposition of NFC through a $\Z_2$ symmetry 
eliminates the possibility of $CP$ breaking in the scalar sector, both explicit and spontaneous. 
In the above mentioned three Higgs doublet model, 
the introduction of a $\Z_2 \times \Z_2$ symmetry guarantees NFC 
while at the same time allowing for either explicit or spontaneous CPV in the scalar sector.
A similar situation was found for 3HDM equipped with the $\Delta(27)$ horizontal symmetry group \cite{geometric},
the minimal model that features the so-called geometric $CP$ violation. 

Recently, a complete analysis has been finalized of all possible discrete symmetry groups 
of the 3HDM scalar sector \cite{classification3HDM} 
and of all possible forms of their spontaneous breaking \cite{breaking-patterns}.
All cases were in line with the generic observation mentioned above: 
if a horizontal symmetry group forbids explicit CPV in the scalar sector, 
then it also automatically prevents spontaneous CPV at the minimum. 
Conversely, if a symmetry group allows for explicit CPV, then
it also allows for spontaneous CPV.

With numerous examples confirming this intriguing observation, one may wonder whether 
this is a truly general phenomenon with no exceptions,
or, on the contrary, it is just a generic trend which can be violated by certain more complicated Higgs sectors.
Even if this phenomenon is not universal, it is worth investigating in which broad classes of models
this conjecture holds.

In this work, we prove that this conjecture holds in multi-Higgs-doublet models 
equipped with abelian (rephasing) horizontal symmetry groups.
There is one minor modification, though, which we will need to make
in order to avoid a somewhat exotic possibility: we pay attention to the explicit $CP$ conservation or violation
in the neutral part of the potential.
With this modification, we show the conjecture to be valid for any multi-Higgs-doublet model
equipped with any rephasing symmetry group.\smallskip

The structure of this paper is as follows. In Section~\ref{section-interplay}, 
we start with a qualitative discussion
of various forms of $CP$ violation originating from the scalar sector.
We illustrate this discussion with several examples from 2HDM and 3HDM.
Then, in Section~\ref{section-method}, we introduce a convenient formalism of treating rephasing transformations,
and use it in Sections~\ref{section-CPC} to prove the conjecture for $CP$-conserving case.
For the $CP$-violating case considered in Section \ref{section-CPV}, the conjecture is also valid 
provided we focus on the neutral scalar sector only.
We also discuss here the exotic possibility of explicit $CP$-violation which arises from the charged Higgs sector.
In Section~\ref{section-conclusions}, we summarize our results.
Finally, Appendix contains a proof of certain mathematical statements mentioned in the main text.

\section{Interplay between symmetries and CP}\label{section-interplay}

\subsection{General remarks}

The generation of $CP$-violation in the scalar sector is linked with the presence of complex parameters 
and complex vacuum expectation values (vevs) of the neutral Higgs fields. 
However, one has to be aware of various subtleties which arise in the presence 
of extra symmetries of the Lagrangian, beyond the gauge symmetry. 
Although many of them have been already discussed in the literature,
we want to dwell on them before venturing into a general analysis.

Let us start with a qualitative discussion of the generation of explicit and spontaneous or $CP$-violation in the scalar sector. 
A multi-Higgs potential contains various terms which link different Higgs fields and
which carry complex coefficients $\lambda_i$. 
Schematically\footnote{This expression can be viewed
as the value of the potential calculated at classical values of neutral components of the 
scalar doublets and omitting the charged ones. 
We deliberately oversimplify the description here to make the main idea as transparent as possible.}, they sum up to 
\be
V \sim \sum_i A_i\cos(\mathrm{phases} + \psi_i)\,,\label{schematic}
\ee
where ``phases'' stands for phases of complex Higgs fields, $\psi_i$ is the phase of $\lambda_i$,
and $A_i$ is a prefactor.
If it happens that all coefficients are real, then all $\psi_i = 0$ or $\pi$, and the potential 
\be
V \sim \sum_i A_i\cos(\mathrm{phases})\,,\label{schematic0}
\ee
becomes symmetric
under ``phases'' $\to$ ``$-$phases'', that is, under the usual $CP$-transformation.

If some of $\psi_i$ do not vanish, then this generic argument does not apply.
However, the presence of nonvanishing $\psi_i$, that is, of complex parameters of the potential,
does not necessarily lead to explicit $CP$-violation.
This can be seen by recalling that in multi-Higgs model with $N$ scalar doublets,
the most general $CP$ transformation that leaves invariant the kinetic part of the lagrangian is 
\be
\phi_i \xrightarrow{CP} U_{ij}\phi_j^*\,,\quad \phi_i^\dagger \xrightarrow{CP} U^*_{ij}\phi_j^T\,,\label{general-CP}
\ee
where $U$ is an arbitrary $N\times N$ unitary matrix acting in the Higgs doublet space.
If we were limited only to the kinetic term, then {\em any} of these transformations could 
play the role of ``the $CP$-transformation'', and in this respect the usual definition, with $U=\mathbb{I}$ 
is not a unique choice.

The Higgs potential does not have to be invariant under all, or even any, of general $CP$ transformations 
of form (\ref{general-CP}). There is explicit $CP$ violation in the scalar sector if and only if
there exists no choice of $U$ which leaves invariant the full scalar lagrangian.
If such a choice exists, then one has $CP$ invariance in spite of some $\psi_i$ being non-zero.
Very schematically, this symmetry, which is usually called generalized $CP$ (gCP) symmetry, 
exploits the compensation of phase changes coming from the usual $CP$ transformation
and the unitary transformation $U$:
\be
\cos(\mathrm{phases} + \psi_i) \xrightarrow{\mathrm{usual\ } CP} \cos(-\mathrm{phases} + \psi_i) 
\xrightarrow{U} \cos(\mathrm{phases} + \psi_i)\,.
\label{schematic2}
\ee
We stress again that the use of general $CP$ transformations is essential to show 
that complex coefficients do not automatically lead to explicit $CP$ violation.
Below, we give a well-known example of such situation, the $A_4$-symmetric 3HDM.

Another useful tool to study the $CP$ properties of the scalar lagrangian consists of Higgs-basis
$CP$-odd invariants \cite{CPV,CP-odd-inv}. Although many examples of these invariants can be written down,
their use in multi-doublet models faces the mathematical challenge of determining 
the complete basis of algebraically independents invariants. 
For 2HDM, this basis was constructed, and invariants indeed turned out to be a useful tool in phenomenological analyses.
Beyond two doublets, this problem has not been solved.

Spontaneous $CP$-violation occurs if there exists a set of (generalized) $CP$ symmetries of the scalar lagrangian 
but none of them leaves invariant the vacuum field configuration \cite{CPV}. 
Using Eq.~(\ref{general-CP}) and the fact that a $CP$-invariant vacuum implies $CP|0\rangle = |0\rangle$,
one derives that, in order for the vacuum of the multi-scalar model to be $CP$ invariant,
the following condition has to be satisfied:
\be
U_{ij} \langle 0|\phi_j|0 \rangle^* = \langle 0|\phi_i|0 \rangle\,.\label{CP-inv-vacuum}
\ee
For real vacua, this condition is trivially satisfied with $U = \mathbb{I}$. However,
it may also be satisfied for complex vacua if the lagrangian contains a gCP symmetry with a non-trivial $U$.

Technically, spontaneous $CP$-violation arises
if several terms of form (\ref{schematic}) or (\ref{schematic2}) are present in the potential.
When minimizing the potential, we do not always have enough freedom to set all cosines to $-1$.
The clash among these requirements can give non-trivial phases to vevs,
so that the vacuum violates the initial $CP$ symmetry.
But once again, this conclusion is not universal: in certain cases, the phases of vevs can take 
special values (also known as calculable phases), 
and, similarly to (\ref{schematic2}), their sign flip can be compensated by an extra transformation,
as shown in (\ref{CP-inv-vacuum}).
In other cases, calculable phases do lead to spontaneous $CP$ violation, 
a phenomenon known as ``geometrical $CP$-violation'' \cite{geometric}.
A further insight into the complexity of the problem was given in \cite{geometric-CPV-larger-N}.
There, the phenomenon of geometrical CPV in models beyond 3HDM was explored
and it was clearly shown that calculable phases do not always lead to spontaneous CPV.

\subsection{Examples from 2HDM and 3HDM}

To illustrate the non-trivial interplay between horizontal symmetries and $CP$ properties,
let us start with the $\Z_2$-symmetric 2HDM, where the symmetry transformation is given by the sign flip of one of the doublets.
The renormalizable potential contains only one term with a complex coefficient: $\lambda_5 (\fd_1\f_2)^2 + h.c.$
One can rephase one doublet to make $\lambda_5$ real, as in (\ref{schematic0}), 
which proves that the model is explicitly $CP$ conserving (CPC). 

Next, suppose that $\lambda_5 >0$. Then, the minimum of the potential requires that there be a relative $\pi/2$
phase between the doublets: $(v_1,\, i v_2)$.
However this vacuum is still not $CP$-violating.
One way to see it is to rephase, before minimization, the second doublet by $\pi/2$: $\phi_2 = i \phi_2'$. 
We would get the same potential but with negative $\lambda_5$, whose minimum is attained at purely real vevs.
The second way of seeing it is to observe that the vacuum $(v_1,\, i v_2)$ is invariant under a gCP transformation,
the usual $CP$ followed by the $\Z_2$ transformation.
Thus, the presence of a gCP symmetry in the vacuum signals the absence of spontaneous $CP$ violation (CPV),
since the criteria (\ref{CP-inv-vacuum}) can be satisfied.

A more intricate situation takes place in 3HDM equipped with various discrete symmetry groups $G$.
Consider first the abelian group $\Z_4$, which is, for 3HDM, the minimal group
preventing both explicit and spontaneous CPV \cite{breaking-patterns}.
In the basis where the generator of $\Z_4$ is $a_4 = \mathrm{diag}(i,\,-i,\,1)$, 
the phase-sensitive part of the Higgs potential contains only two terms:
\be
V_1(\Z_4) = \lambda_1 (\fd_1 \f_2)^2 + \lambda_2 (\fd_1 \f_3)(\fd_2 \f_3) + h.c.\label{V1-Z4}
\ee
We have two complex coefficients and we can rephase two doublets with respect to the third. 
This freedom is sufficient to make both coefficients real,
which proves that $\Z_4$ automatically leads to explicit CPC.
This example makes it clear that if there is a sufficiently small number of phase sensitive terms,
one can use the rephasing freedom to arrive at (\ref{schematic0}).

Proving that $\Z_4$ prevents spontaneous CPV is less trivial \cite{breaking-patterns}.
One can substitute a generic vev alignment $\lr{\phi_i^0} = v_i e^{i\xi_i}/\sqrt{2}$ into (\ref{V1-Z4}),
differentiate it with respect to phases, and check the consequences 
for different assumptions about the number of zero vevs.
\begin{itemize}
\item
If all $v_i \not = 0$, then $\xi_1$ and $\xi_2$ are opposite and are multiples of $\pi/4$,
which implies existence of a gCP symmetry of the vacuum.
A similar situation takes place for $v_3 = 0$.
\item
If $v_1$ or $v_2=0$, then the relative phase of the remaining two vevs can be arbitrary, 
which formally constitutes a $CP$-violating solution,
but one can show that it is a saddle point, not a minimum.
\item
If two vevs are zero, the residual phase is unphysical and can be set to zero by a $U(1)$ transformation.
\end{itemize}
Once again, the counting of terms plays an important role.
If we have a sufficiently small number of terms, then we can find the phases which set all of the non-zero terms to $-1$.
This leads to calculable phases which inherit some residual symmetry.

Now consider 3HDM with three phase-sensitive terms, so that one cannot make all coefficients real.
Even in this case there is room for explicit CPC as in (\ref{schematic2}).
For example, in the $A_4$-symmetric 3HDM, the phase sensitive part takes form
\be
V_1(A_4) = \lambda \left[(\fd_1 \f_2)^2 + (\fd_2 \f_3)^2 + (\fd_3 \f_1)^2\right] + h.c.\label{V1-A4}
\ee
with complex $\lambda$. This potential is invariant under sign flips of individual
doublets and under their cyclic permutations. 
The model is explicitly $CP$-conserving, with the gCP symmetries being usual $CP$ followed 
by an exchange of any pair of doublets, which provides an example of the situation (\ref{schematic2}).
Also, when minimizing the $A_4$ potential, one can have complex vevs but still
there always remains a gCP symmetry respecting the vacuum alignment. Thus, $A_4$ prevents both explicit
and spontaneous $CP$ violation.

One might suspect that this example lacks CPV because the potential contains only one overall complex coefficient. 
Consider another similarly looking situation, $\Delta(54)$-symmetric 3HDM. Its phase-sensitive potential is
\be
V_1(\Delta(54)) = \lambda \left[(\fd_1 \f_2)(\fd_1\f_3) + (\fd_2 \f_3)(\fd_2\f_1) + (\fd_3 \f_1)(\fd_3\f_2)\right] + h.c.\label{V1-Delta54}
\ee
with complex $\lambda$. It is symmetric under any permutations of doublets and under 
the order-3 rephasing diag$(1,\,\omega,\,\omega^2)$, where $\omega \equiv \exp(2\pi/3)$,
which form the group $\Delta(54)$. This model does not have any
gCP symmetry and is, therefore, explicitly CPV.
Its CPC version allows for the spontaneous CPV, which is known since 1984 as the geometrical CPV \cite{geometric}.
Thus, $\Delta(54)$ allows for either explicit or spontaneous $CP$ violation.

In order to clarify the origin of so distinct $CP$-properties in similarly looking models,
let us introduce a convenient notation. Take the $i$-th term from the phase-sensitive part 
of the potential $V_1$ and rephase all doublets by their own $\alpha_j$, $j = 1,\dots, N$.
The term picks up an extra phase which can be generically written as $\sum_j d_{ij}\alpha_j$. 
The integer-valued matrix $d_{ij}$ can be easily written for any model just by looking at the potential.
In particular, for the two cases of 3HDM considered above, we have
\be
d(A_4) = \left(
\begin{array}{rrrr}
-2 & 2 & 0\\
0 & -2 & 2\\
2 & 0 & -2
\end{array}
\right)\,, \quad 
d(\Delta(54)) = \left(
\begin{array}{rrrr}
-2 & 1 & 1\\
1 & -2 & 1\\
1 & 1 & -2
\end{array}
\right)\,.\label{dA4D54}
\ee
These matrices make it clear why the two models are so different in their $CP$ consequences.
The matrix $d(A_4)$ enjoys the following property: if we flip the overall sign, then, by an appropriate permutations
of columns and rows, we can recover the original matrix.
It is this property that enables the transformation (\ref{schematic2}).
The matrix $d(\Delta(54))$ does not have this feature: $-d$ is essentially different from $d$ and cannot be recovered
by permutations. Therefore, transformation (\ref{schematic2}) is impossible.

As for the spontaneous CPV, we are not aware of an equally simple argument.
We can only state that the exact minimization of $A_4$ and $\Delta(54)$ potentials
leads to vevs with very rigid structures \cite{breaking-patterns}.
In the case of $A_4$, this rigid structure inherits a gCP, while in the $\Delta(54)$
case, the vev alignment, despite being rigid, allows for spontaneous CPV.

These and all other examples in 2HDM and 3HDM, despite various intricacies, are all consistent
with the observation that CPV comes in pairs, meaning that
if the horizontal $G$ prevents explicit CPV, then it also prevents spontaneous CPV.
If $G$ allows explicit CPV, then, in its explicitly $CP$ conserving version,
it allows for spontaneous CPV.
Intrigued by this seemingly robust correlation, we start a systematic investigation
of how general it is.
This paper is devoted to the broad class of NHDM models in which $G$ is an abelian group,
represented by rephasing. More elaborate cases with non-abelian groups, 
represented by rephasing and permutations, are postponed for future study.

\section{Rephasing symmetries and CPV}\label{section-method}

We start with a brief reminder of how rephasing symmetries of the scalar sector in $N$-Higgs-doublet model (NHDM)
can be efficiently analyzed \cite{abelianNHDM}.
The Higgs potential is split into phase-independent part $V_0$, which is symmetric under $[U(1)]^N$ including
the overall rephasing, and the phase-sensitive part $V_1$.
The latter contains $m$ terms with complex coefficients $\lambda_i$, $i = 1, \dots, m$, as well as their conjugates.
Next, we evaluate $V_1$ at quasiclassical values $\lr{\phi_i^0} = v_i e^{i\xi_i}/\sqrt{2}$;
for example,  
\be
\lambda_1 (\fd_1 \f_2)(\fd_1\f_3) + h.c. \to {1\over 2}|\lambda_1| v_1^2 v_2 v_3 \cos(-2\xi_1 + \xi_2 + \xi_3 + \psi_1)\,,
\ee
where $\psi_1$ is the phase of $\lambda_1$.
Then we write the argument of the cosine of the $i$-th term as $d_{ij} \xi_j + \psi_i$,
where the integer-valued matrix $d_{ij}$ picks up the phases for each term. In the above example,
$d_{1j} = (-2,\, 1,\, 1,\, 0,\, \dots,\, 0)$. Note also that the powers of $v_i$ in front of the cosine can also be written
as $\prod_j v_j^{|d|_{ij}}$, where entries of the matrix $|d|_{ij}$ are just absolute values of the corresponding entries of 
$d_{ij}$.
With this notation, $V_1$ takes at the quasiclassical values of the Higgs fields the following compact form:
\be
V_1 = \sum_{i=1}^m |\lambda_i| \left( \prod_{j=1}^N v_j^{|d_{ij}|}\right) \cos(d_{ij}\xi_j + \psi_i)\,.\label{V1-vev}
\ee
Rephasing symmetries are such shifts of phases $\xi_j$ which leave each term in (\ref{V1-vev}) invariant.
They arise as the solutions of the system 
\be
d_{ij}\alpha_j = 2\pi n_i\,,\label{rephasing-symmetries}
\ee
with any integer $n_i$.
The full rephasing symmetry group of a model is given by all solutions of this system.

Solving this system, reconstructing the group, and analyzing its solutions is efficiently done with
the Smith normal form technique developed in \cite{abelianNHDM} for the scalar sector
and in \cite{IN2013} for Yukawa sector in NHDM. Here, we do not need this technique, apart from
the proof of a technical statement given in the Appendix.
However, we will exploit the following important properties of the matrix $d_{ij}$.
\begin{itemize}
\item
The matrix $d_{ij}$ is a rectangular $m \times N$ matrix ($m$ rows, $N$ columns).
The value of $m$ can be larger or smaller than $N$, but it is always true that rank $d \le N-1$,
because, in each row, the sum of all entries must be zero (the numbers of $\phi$ and $\phi^\dagger$
are equal in each term). The consequence is that there is always a solution of (\ref{rephasing-symmetries}) in the form of 
$\alpha_j = \alpha (1,\,1,\,\dots,\,1)$ with arbitrary $\alpha$. 
These solutions form the $U(1)$ group of overall rephasing, 
which is a part of the hypercharge symmetry group and is always present for any Higgs potential. 
When we impose rephasing symmetries, we mean extra solutions of (\ref{rephasing-symmetries}) in addition to
this trivial one. 
\item
If rank $d < N-1$, there exist other solutions to $d_{ij}\alpha_j =0$, that is, other subspaces annihilated by $d_{ij}$.
They lead to continuous symmetries of the model. These symmetries either remain unbroken after EWSB,
or, when broken, they lead to massless scalars. In either case, the situation is not related to the problem we consider.
Thus, we are interested in models with rank $d = N-1$.
\end{itemize}
Now, there are several options for $m$.
\begin{itemize}
\item
$m < N-1$ leads to rank$\,d < N-1$, and we disregard this case.
\item
$m = N-1$ will constitute our main $CP$-conserving case.
The rectangular $(N-1)\times N$ matrix $d$ is then a full rank matrix,
which means that all its rows are linearly independent. 
In addition, by removing a column, one can arrive at a square $(N-1)\times(N-1)$ matrix $d'$
which is invertible.
\item
$m = N$ will constitute our main $CP$-violating case. 
The $N$ rows of the matrix are now linearly dependent: 
that is, there exist coefficients
$c_i$ not all being zeros such that $c_i d_{ij} = 0$ for all $j$.
\item
$m > N$, which makes the model even more $CP$-violating, with extra complex
free parameters and extra possibility to get non-zero phases. 
This situation does not give anything new with respect to the previous case.
\end{itemize}

In the recent work \cite{Haber-Surujon}, the conditions for the spontaneous CPV
were derived within a spurion formalism, which is close in spirit to ours.
Each phase-sensitive term of the potential comes with its coefficient, which plays the role of a spurion;
thus, the number of spurions $N_s$ is equal to our $m$.
The number of linearly independent charge vectors denoted in \cite{Haber-Surujon} by $r$
corresponds to our $N-1$. The necessary condition for spontaneous CPV derived in \cite{Haber-Surujon}
is $N_s > r$, which is compatible with our classification.
We think, however, that the formalism we use, based on left and right spaces of $d_{ij}$, 
is more elegant and allows us to reach conclusions about the interplay between rephasing symmetries and various forms of CPV.

\section{$CP$-conserving case}\label{section-CPC}

\subsection{Generic arguments}
Suppose that the rephasing group $G$ is such that it allows exactly $m = N-1$ different terms in $V_1$,
and that rank$\,d = N-1$. Then, the model is explicitly $CP$-conserving.
Indeed, start with cosine arguments $d_{ij}\xi_j+\psi_i$ and rephase the doublets 
$\xi_j \mapsto \xi_j + \delta_j$, keeping one doublet (which can be labeled to be the last one) unchanged: $\delta_N = 0$.
The arguments of the cosines change to $d_{ij}\xi_j + d_{ij}'\delta_j + \psi_i$,
where $d_{ij}'$ is the square $m\times m$ matrix obtained from $d$ by removing the last column.
Since $d'$ is invertible, the system $d_{ij}'\delta_j + \psi_i = 0$ always has a solution.
Thus, one can always rephase the doublets to make all coefficients real.

Next, we want to prove that, in this case, there is also no room for spontaneous $CP$-violation. 
Let us introduce the shorthand notation
\be
V_1 = \sum_{i=1}^{m} A_i \cos(d_{ij}\xi_j)\,,\quad A_i \equiv |\lambda_i| \prod_{j=1}^N v_j^{|d_{ij}|}\,.
\ee
Differentiating $V_1$ with respect to $\xi_j$, $j = 1, \dots, N$, we get the phase stationarity condition,
which is equivalent to setting to zero the following 1-form:
\be
d_\xi V \equiv - (A_1 s_1, \, \dots,\, A_m s_m) \cdot d \cdot \triplet{d\xi_1}{\vdots}{d\xi_N} = 0\,,
\label{1-form}
\ee
where $s_i \equiv \sin(d_{ij}\xi_j)$. Since it holds for all directions $d\xi_j$, we get
\be
\sum_i A_i s_i \, d_{ij} = 0 \quad \forall\  j = 1,\, \dots,\, N\,. \label{1-form-b}
\ee
Since matrix $d_{ij}$ has rank $m$, its $m$ rows are linearly independent. 
Therefore, if a linear combination of these rows is zero, as written in (\ref{1-form-b}), 
then all coefficients must be zero:
\be
A_i s_i = 0 \quad \forall\  i = 1,\, \dots,\, m\,. \label{aisi-CPC}
\ee

Next, the analysis splits into two cases: if all $v_i \not = 0$, or if some $v_i = 0$.
In the former case, each $A_i\not = 0$, and (\ref{aisi-CPC}) simply means that all $s_i=0$.
Then, we obtain 
\bea
&&d_{ij}\xi_j = 0\ \mbox{or}\ \pi \mod 2\pi\quad \quad \forall\  i = 1,\, \dots,\, m\,,\nonumber\\
&\Leftrightarrow & d_{ij}\xi_j = - d_{ij}\xi_j  \mod 2\pi\,,  \nonumber\\
&\Leftrightarrow & \xi_i = - \xi_i + \alpha_i\,,\label{gcp-condition}
\eea
where $\alpha_i$ is a solution of the system $d_{ij}\alpha_j = 0  \mod 2\pi$.
But we know that all solutions of this system form the rephasing symmetry group, see (\ref{rephasing-symmetries}).
Therefore, phases of vevs $\xi_i$ satisfy (\ref{gcp-condition}) with $\alpha_i$ being
a symmetry transformation. As a result, vevs are invariant under some gCP: 
$\lr{\phi_j} \mapsto e^{i\alpha_j}\lr{\phi_j}^* = \lr{\phi_j}$,
and no spontaneous CPV occurs.

If some vevs are zero, the analysis becomes more complicated. 
The presence of zero vevs means that some among $m$ conditions (\ref{aisi-CPC}) are satisfied
by $A_i = 0$, which places no restriction on the sines.
However one can exploit the fact that the matrix $d$ affects both $A_i$ and $s_i$,
namely, if $v_j$ is present in some $A_i$,
then its phase $\xi_j$ is present in $s_i$.
Then, one can extract from $d_{ij}$ a submatrix $\tilde d_{ij}$, which corresponds 
only to those conditions for which $s_i = 0$ and which couple only to doublets with non-zero vevs. 
Note that all rows of $\tilde d$ are also linearly independent.
The phases of these non-zero vevs then satisfy 
\be
\tilde d_{ij} \xi_j = - \tilde d_{ij} \xi_j  \mod 2\pi \,, 
\ee
which is analogous to (\ref{gcp-condition}) but has fewer conditions and fewer phases involved.
In the Appendix we show that two possibilities can take place. Either there are too few conditions on sines,
and in this case we either have massless scalars or a saddle point,
or the number of conditions is just right to fix phases $\xi_j$, but then there remains a residual rephasing symmetry
with angles $\alpha_j$, which satisfy $\xi_j = - \xi_j + \alpha_j$.

The overall conclusion is that whatever values $v_i$ take,
a minimum always contains a residual generalized $CP$ symmetry.
Thus, spontaneous $CP$ violation cannot take place.

\subsection{Illustrations}

In the simplest example of this kind, $\Z_2$-symmetric 2HDM, 
$V_1 = \lambda_5 (\fd_1\f_2)^2 + h.c.$ The matrix $d_{ij} = (-2,\,2)$ has $m=1$ row and $N=2$ columns,
and the solution of $d_{ij}\alpha_j = 2\pi n_i$ gives $\alpha_j = (0, \pi)$, up to overall rephasing.
Clearly, the model is explicitly $CP$-conserving because we are able to rotate away the phase of the complex parameter $\lambda_5$.
If we look for a minimum with non-zero vevs, we set $\sin(2\xi_2-2\xi_1) =0$, from which we can obtain
real solutions $(v_1,\, \pm v_2)$ and complex solutions $(v_1,\, \pm i v_2)$,
which still respect a gCP symmetry.

In the more elaborate case of $\Z_4$-symmetric 3HDM with the potential (\ref{V1-Z4}),
the matrix $d$ is 
\be
d = \left(
\begin{array}{ccc}
-2 & 2 & 0 \\
-1 & -1 & 2
\end{array}
\right)\,.
\ee
Explicit differentiation and solution of the system of equations shows that, if all vevs are non-zero, 
one can obtain complex vevs of the following types \cite{breaking-patterns}
\be
(\pm v_1 e^{i\pi/4},\, \pm v_2 e^{-i\pi/4},\, v_3)\,,\quad (\pm i v_1,\, \mp i v_2,\, v_3)\,. 
\ee
In any case, complex conjugation can be compensated by applying a transformation from the $\Z_4$ symmetry group.
If one assumes, instead, that $v_1=0$, then the phase conditions disappear and one gets $(0, v_2  e^{i\xi_2},\, v_3)$
as a viable solution of the phase stationarity condition.
However in this case $\partial^2 V/\partial \xi_2^2 = 0$ and $\partial^2 V/\partial \xi_2 \partial v_1 \not = 0$;
therefore, this vev alignment corresponds to a saddle point \cite{breaking-patterns}.
One can also check other zero vev alignments and observe that in all cases
the model is $CP$ conserving, both explicitly and spontaneously.

\section{$CP$-violating case}\label{section-CPV}

\subsection{Generic arguments}

Suppose now that a rephasing symmetry group allows for exactly $m=N$ phase-sensitive terms
in the Higgs potential. The matrix $d_{ij}$ is then a square $N\times N$ matrix with rank $N-1$.
If one starts with (\ref{V1-vev}) with arbitrary phases $\psi_i$,
then it will be impossible to rephase them away.
Indeed, for that one would need to solve system $d_{ij}\xi_j = -\psi_i$
for $\xi_i$. But since $d$ is not invertible, this is impossible for generic $\psi_i$.
One can, however, make all $\psi_i$ equal.
Thus, this symmetry group is compatible with explicit $CP$ violation in the Higgs sector.

Now, consider its explicitly $CP$ conserving version, in which all $\psi_i$ are set to zero.
Differentiating $V_1$ with respect to phases still leads to (\ref{1-form-b}). 
But now, in addition to the previous solution with all $A_i s_i =0$, 
we can have a non-trivial solution with not all $A_i$ and $s_i$ being zero (we call it a non-zero solution).
This solution is unique up to an overall factor because the matrix $d$ has a one-dimensional
kernel.

Now, the mere fact that there exists a stationary point with non-zero $A_i s_i$
automatically makes it $CP$-violating.
Indeed, suppose there is a gCP transformation based on a rephasing symmetry $\alpha_j$ which is still preserved at the minimum.
It acts on vev phases by $\xi_i \mapsto -\xi_i + \alpha_i = \xi_i$, which means 
$d_{ij}\xi_j \mapsto -d_{ij}\xi_j = d_{ij}\xi_j$.
This can happen only when all $d_{ij}\xi_j = 0$ or $\pi$, so that all $s_i = 0$, 
which contradicts the definition of a non-zero solution. 
Therefore, in such a solution, no residual gCP symmetry exists, and we obtain spontaneous $CP$ violation.

\subsection{Illustration}

Again, to give an illustration, consider the original Weinberg's model \cite{Weinberg}, 
the $\Z_2\times \Z_2$ 3HDM symmetric under sign flips of individual
doublets. The phase-sensitive part of the potential is 
\be
V_1(\Z_2\times \Z_2) = \lambda_1 (\fd_1 \f_2)^2 + \lambda_2 (\fd_2 \f_3)^2 + \lambda_3 (\fd_3 \f_1)^2 + h.c.\label{V1-Z2Z2}
\ee
with complex $\lambda_i$. The matrix $d$ takes the same form as in the $A_4$ case (\ref{dA4D54}),
which is not surprising because $A_4$ is just an extension of $\Z_2\times \Z_2$ 
by the cyclic permutation group $\Z_3$.
The model now contains three complex parameters $\lambda_i$ which cannot be made simultaneously real,
and, unlike the $A_4$ case, one cannot use permutations and resort to (\ref{schematic2}).
Thus, the model is explicitly CPV.

In its explicitly CPC version \cite{Branco}, we proceed with minimization and have a non-zero solution of (\ref{1-form-b}) in the form
\be
A_1 s_1 = A_2 s_2 = A_3 s_3\,.\label{conditionZ2Z2}
\ee
For a region of vevs and $\lambda$'s which satisfy certain triangle inequalities \cite{Branco,CPV}, 
this solution does exist, and it displays spontaneous $CP$ violation.

\subsection{A peculiar source of $CP$-violation}

In the above derivation, we showed that if a non-zero solution to (\ref{1-form-b}) exists,
then it is spontaneously $CP$-violating.
Does such a solution always exist, at least for some parameters of the model?
The main line of arguments proceeds as follows.
Solving (\ref{1-form-b}) leads to certain relations among $A_i s_i$, for example, (\ref{conditionZ2Z2}).
Let us pick up generic phases and generic vevs; then one can adjust the absolute values
of the coefficients $\lambda_i$ in each $A_i$ term to fulfill these conditions.
Thus, the chosen vev alignment gives the desired non-zero CPV solution for the model with adjusted coefficients.

However, there remains one peculiar possibility in which this main argument fails.
Considering it in detail below, we arrive at a novel possible way the $CP$ violation can be generated by the scalar potential.

Consider a model with four doublets and with the following phase-sensitive terms
\bea
V_1 &=& \lambda_1 (\fd_2 \f_1)(\fd_3 \f_1) + \lambda_2 (\fd_1 \f_4)(\fd_2 \f_4) +
\lambda_3 (\fd_1 \f_3)(\fd_4 \f_3) + \lambda_4 (\fd_3 \f_2)(\fd_4 \f_2)\nonumber\\
&&+ \lambda_5 (\fd_1 \f_2)(\fd_3 \f_4) + \lambda_6 (\fd_1 \f_4)(\fd_3 \f_2) + h.c.\label{hybrid}
\eea
with complex parameters $\lambda_i$. 
This model has the symmetry group $\Z_5$ generated by the following transformation:
\be
a_5 = \mathrm{diag}\left(\eta^2,\,1,\,\eta^4,\,\eta\right)\,,\quad \mbox{where}\quad \eta^5 = 1\,.
\ee
The potential (\ref{hybrid}) contains all renormalizable terms consistent with this symmetry.
The matrix $d_{ij}$ is:
\be
d = \left(
\begin{array}{rrrr}
2 & -1 & -1 & 0\\
-1 & -1 & 0 & 2\\
-1 & 0 & 2 & -1 \\
0 & 2 & -1 & -1 \\
-1 & 1 & -1 & 1\\
-1 & 1 & -1 & 1
\end{array}
\right)\,,\quad \mathrm{rank}\,d = 3\,.
\ee
Note that the last two rows are identical because the corresponding terms 
transform in exactly the same way under rephasing of doublets.
This model is clearly of the $CP$-violating class due to a large number of complex coefficients.
In its $CP$ conserving version, with real $\lambda_i$, 
one can apply the results of the previous subsection
and find spontaneously $CP$-violating minimum.
One can construct several different non-zero solution of (\ref{1-form-b}):
\bs
\bea
\mbox{solution 1}: && A_1 s_1 = A_2 s_2 = A_3 s_3 = A_4 s_4 = 0\,,\quad A_5 s_5 = - A_6 s_6 \not = 0\,,\label{solution1} \\
\mbox{solution 2}: && A_1 s_1 = A_2 s_2 = A_3 s_3 = A_4 s_4 \not = 0\,,\quad A_5 s_5 = A_6 s_6 = 0\,,\label{solution2} \\
\mbox{solution 3}: && A_1 s_1 = A_3 s_3 = A_5 s_5 \not = 0\,,\quad A_2 s_2  = A_4 s_4 = - A_6 s_6 \not = 0\,,\label{solution3}
\eea 
\es
and their linear combinations.
However, we know that, by construction, $s_5 = s_6$, and that $A_5$ and $A_6$ contain exactly the same vev
combinations.
It is possible to satisfy $A_5 s_5 = - A_6 s_6 \not = 0$ only if $\lambda_5 = - \lambda_6$.
Thus, a {\em generic} $\Z_5$ symmetric 4HDM does not allow for solution 1 given by (\ref{solution1}).
Luckily, other solutions exist, so that there remains a possibility for spontaneous CPV.
Thus, this model, too, complies with the main observation.

Imagine now that we truncated the potential (\ref{hybrid}) by leaving out the $\lambda_3$ and $\lambda_4$ terms.
Then, we would have a 4HDM model with four phase-sensitive terms. 
By applying the general results of the previous subsection, 
we deduce the possibility for explicit $CP$-violation. 
But there will be no room for spontaneous $CP$ violation because
solutions 2 and 3 given in (\ref{solution2}) and (\ref{solution3}) are unavailable.
In this case we would obtain a counterexample to the general trend: 
a symmetry-driven model with explicit CPV but no spontaneous CPV.
However, this explicit CPV would be of a peculiar kind.
The truncated potential can be rewritten as
\bea
V_1 &=& \lambda_1 (\fd_2 \f_1)(\fd_3 \f_1) + \lambda_2 (\fd_1 \f_4)(\fd_2 \f_4)
+ \lambda'_5 (\fd_1 \f_2)(\fd_3 \f_4) \nonumber\\
&& + \lambda_6 \left[(\fd_1 \f_4)(\fd_3 \f_2) - (\fd_1 \f_2)(\fd_3 \f_4)\right]+ h.c.\label{hybrid2}
\eea
By rephasing doublets, one can make $\lambda_1$, $\lambda_2$, $\lambda'_5$ real,
while $\lambda_6$ stays complex. But this complex coefficient stands in front of an expression with
{\em zero vacuum expectation value}. This term introduces explicitly CPV effects only through
the charged Higgs sector and never through the neutral one, at least at the tree-level.
If one focuses on the neutral Higgs sector exclusively, the $\lambda_6$ term in (\ref{hybrid2}) is absent. 
The matrix $d$ has one row less, it falls into the CPC case considered earlier,
and the absence of spontaneous CPC is then compatible with the general arguments.\smallskip

We stress that the example considered in this subsection, with the full $\Z_5$-symmetric potential
(\ref{hybrid}), is not a counterexample to our general observation.
We tried to construct a counterexample realizing the above idea with even more Higgs doublets,
but we could not find any.
It remains to be checked whether a such an example can be constructed at all. 

\section{Discussion and conclusions}\label{section-conclusions}

It has been known for a long time that imposing an extra horizontal symmetry in multi-Higgs models
can affect $CP$-violation (CPV) coming from the scalar sector.
It was also noticed that in all known examples the following correlation is valid:
if a symmetry group
prevents explicit CPV, then it also prevents spontaneous CPV, 
and if a symmetry group allows for explicit CPV, then it also allows for spontaneous CPV
in the explicitly $CP$-conserving version.
All known examples in 2HDM and 3HDM go along with this observation, so it is natural to ask 
how general this feature is.

In this work, we started to systematically investigate this old and intriguing observation.
We proved this trend to be valid for abelian groups and any number of doublets.
Our formalism also offers a more transparent view why various forms of CPV are present or absent 
in specific models considered in literature.

When investigating the $CP$-violating part of the observation, we noticed a peculiar possibility
of explicit generation of $CP$-violation via the charged Higgs sector.
Such a model would possess ``hybrid'' $CP$-properties:
despite being explicitly $CP$-violating, it remains explicitly $CP$-conserving within
the neutral Higgs sector alone and it does not allow for spontaneous CPV.
Although we have not succeeded in building a full model that realizes this idea, 
we leave it as an open possibility.

If such an example is found, constituting a counterexample to the original general observation,
we may slightly modify the conjecture to eliminate this clash. The corrected conjecture reads:
if a symmetry group prevents explicit CPV {\em in the neutral Higgs sector},
then it also prevents spontaneous CPV; 
if a group allows for explicit CPV in the neutral sector, then it also allows for spontaneous CPV.
In this work we proved this modified conjecture for any abelian (rephasing) symmetry group.
Whether it holds for non-abelian groups such as permutation groups, remains to be investigated.\bigskip

We thank Celso Nishi for useful comments. 
This work was supported by the Portuguese
\textit{Fun\-da\-\c{c}\~{a}o para a Ci\^{e}ncia e a Tecnologia} (FCT)
under contracts UID/FIS/00777/2013 and CERN/FIS-NUC/0010/2015,
which are partially funded through POCTI (FEDER),
COMPETE,
QREN
and the EU.
I.P.I. acknowledges funding from the \textit{Funda\c{c}\~{a}o para a Ci\^{e}ncia e Tecnologia}
through the FCT Investigator contract IF/00989/2014/CP1214/CT0004
under the IF2014 Programme.

\appendix

\section{Symmetries of a submatrix}

Here, we give a detailed proof that, in the explicitly $CP$-conserving model protected by a rephasing symmetry group, 
there remains a generalized $CP$ symmetry even if some vevs are zero.

We are free to choose which vevs we want to set to zero; the final conclusion should not depend on this choice.
Let us denote the number of such zero vevs by $n_0$.
We rearrange the doublets so that the zero vevs come at the end:
\be
v_i = (v_1,\, v_2,\, \dots,\ v_{N-n_0},\,0,\,\dots,\,0)\,.
\ee
We will collectively call the doublets with zero vevs the ``inert space''.

We exploit the fact that matrix $d$ affects both $A_i$ and $s_i$: if $v_j$ is present in some $A_i$,
then its phase $\xi_j$ is present in $s_i$.
The presence of zero vevs means that some among $m$ conditions (\ref{aisi-CPC}) are satisfied
by $A_i = 0$, which places no restriction on the sines.
Let us rearrange $m$ conditions $A_is_i=0$ in such a way that the first $m_s$ conditions 
have non-zero $A_i$ and, therefore, are satisfied by $s_i=0$,
while the remaining $m_a$ conditions are coupled to the inert space and are satisfied by $A_i = 0$,
placing no restriction on $s_i$.
Then, the matrix $d$ takes the block form
\be
d = 
\left (
\begin{array}{c|c}
\tilde{d} & 0 \\ \hline
B & C \end{array}
\right )\,,\label{d-block}
\ee
The first $N-n_0$ columns of $d$, which form the matrices $\tilde d$ and $B$,
are coupled to the non-inert space, while the last $n_0$ columns are coupled to the inert space.
The phases of the non-inert vevs are defined by $m_s$ conditions $s_i \equiv \sin(\tilde d_{ij}\xi_j) = 0$, 
which translates to
\be
\tilde d_{ij} \xi_j = - \tilde d_{ij} \xi_j \mod 2\pi \,, 
\ee
with $i = 1,\,\dots,\,m_s$ and $j = 1,\,\dots,\,N-n_0$.
This is analogous to (\ref{gcp-condition}) but has fewer conditions and fewer phases involved.
Note also that all rows of $\tilde d$ are linearly independent.

Numbers $m_s$ and $n_0$ are, in principle, independent: they reflect our arbitrary choice of 
which inert space we want to test. However, exploring matrix $\tilde d$, we arrive at a similar classification
of cases as before.
\begin{itemize}
\item
If $m_s < N-n_0 - 1$, which includes the case of $m_s = 0$ (recall the $\Z_4$ 3HDM example with $v_2=0$),
then there are too few conditions to fix the remaining non-inert phases.
Then, some phases of non-inert vevs, $\xi_k$, remain free parameters, and we obtain a continuum of stationary points.
Second derivatives along these directions are also zero: $\partial^2 V/\partial \xi_k^2 = 0$.
Depending on whether the off-diagonal hessian elements such as $\partial^2 V/\partial \xi_k \partial v_i$, 
where $v_i$ is from the inert space, are zero or not, we obtain either massless physical scalars or a saddle point.
Either situation is non-physical.
\item
If $m_s = N-n_0 - 1$, then $\tilde d$ inherits from $d$ the property of being full rank and having only one more column than rows.
Since the overall phase of all doublets can be changed at will,
we fix the phase of one of the vevs, which we label to be the first one, and consider the reduced matrix $\tilde d'$ without the first column.
Then, $\tilde d'$ is a square invertible matrix.
Just like in the previous subsection,
we can write 
\be
\tilde d_{ij}' \xi_j = - \tilde d_{ij}' \xi_j \quad \Leftrightarrow \quad \xi_j = - \xi_j + \tilde \alpha_j\,,\quad j = 1,\,\dots N-n_0\,,
\label{tilde-d}
\ee
where $\tilde \alpha_j$ is a rephasing symmetry of first $m_s$ terms. We show 
below that $\tilde \alpha_j$ always corresponds to a symmetry of the full matrix $d$, if restricted to the non-inert space. 
Therefore, in this case, too, we have a residual gCP symmetry.
\item
Finally, situation $m_s > N-n_0 -1$ is impossible. We would have $m_s$ rows which,
on the one hand, are vectors in space with dimensionality less than $m_s$, but on the other hand, must be
linearly independent. 
\end{itemize}
The overall conclusion is that whatever are the values $v_i$,
a minimum always contains a residual generalized $CP$ symmetry.
Thus, spontaneous $CP$ violation cannot take place.\smallskip

It remains to be proven that solution (\ref{tilde-d}) always produces a symmetry of the full matrix $d$.
Start again from the block form (\ref{d-block}) and, by applying elementary steps on columns and rows 
(sign flip, exchanges, and addition), bring the matrices $\tilde d$ and $C$ to their Smith normal forms,
see details in \cite{abelianNHDM,IN2013}.  
In this procedure, we do not mix inert and non-inert spaces, and do not mix conditions $s_i=0$ with conditions $A_i=0$.
The matrix $d$ then takes the partially diagonalized form
\be
d = 
\left (
\begin{array}{cccc|ccc}
0 & d_1 & & & & & \\ 
 &  & \ddots & & & & \\ 
& & & d_{m_s}& & & \\ \hline
 & & & &c_1 & & \\ 
 & & B & & & \ddots& \\ 
 & & & & & & c_{n_0} \\ 
\end{array}
\right )\,,
\ee
with non-zero diagonal entries $d_i$ and with a generic matrix $B$.
The system $\tilde d_{ij} \tilde \alpha_j = 2\pi \tilde n_i$ is decoupled, and
its solutions are arbitrary sums of $\tilde \alpha_j = 2\pi/d_j$ with integer coefficients for the non-inert
doublets and {\em arbitrary} $\alpha_j$ for inert doublets.
Thus, each discrete solution of $\tilde d_{ij} \tilde \alpha_j = 2\pi \tilde n_i$ comes with the $n_0$-dimensional
torus of solutions with arbitrary phases in the inert space.
For example, $\alpha_j$ of the form
\be
\alpha_j = \left(
\begin{array}{c}
0 \\
\vdots \\
0 \\
{2\pi n_{m_s}\over d_{m_s}} \\
\beta_1 \\
\vdots\\
\beta_{n_0}
\end{array}
\right)\label{alpha-j-torus}
\ee
is a solution of the upper half of the system $d_{ij} \alpha_j = 2\pi n_i$ for any $\beta_1,\, \dots,\, \beta_{n_0}$.

Now, in order to extend it to the lower half, we can adjust $n_0$ parameters $\beta_k$ to satisfy $n_0$
conditions $B_{k m_s}\cdot 2\pi/d_{m_s} + c_k \beta_k = 0$.
As the result, for each solution $\tilde d_{ij} \tilde \alpha_j = 2\pi \tilde n_i$, we get a {\em single} solution 
of $d_{ij} \alpha_j = 2\pi n_i$, and these two solutions coincide in the non-inert space.
Different solutions do not need to form a closed group within the inert space; 
they are just required to agree with the symmetry transformations
in the non-inert space.
Thus, existence of solution (\ref{tilde-d}) always implies a residual gCP transformation in the vacuum
belonging to the original symmetry group.

\end{document}